\documentclass[aps,preprint,showpacs,floatfix]{revtex4-1}
\usepackage{graphicx,bm,amsmath,dcolumn}
 \usepackage{geometry}
\begin{document}
\newcommand{\half}{\frac{1}{2}}
\newcommand{\ith}{^{(i)}}
\newcommand{\im}{^{(i-1)}}
\newcommand{\gae}
{\,\hbox{\lower0.5ex\hbox{$\sim$}\llap{\raise0.5ex\hbox{$>$}}}\,}
\newcommand{\lae}
{\,\hbox{\lower0.5ex\hbox{$\sim$}\llap{\raise0.5ex\hbox{$<$}}}\,}
\newcommand{\be}{\begin{equation}}
\newcommand{\ee}{\end{equation}}
\newcommand{\bea}{\begin{eqnarray}}
\newcommand{\eea}{\end{eqnarray}}

\title{Potts and percolation models on bowtie lattices}
\author{Chengxiang Ding$^{1}$, Yancheng Wang$^{2}$ and Yang Li$^3$}
\affiliation{$^{1}$Department of Applied Physics, Anhui University of Technology, Maanshan 243002, P. R. China }
\affiliation{$^{2}$Physics Department, Beijing Normal University, Beijing 100875, P. R. China }
\affiliation{$^{3}$Electrical Engineering Department, Tianjing agricultural University, Tianjing 300384,  P. R. China }
\date{\today} 
\begin{abstract}
We give the exact critical frontier of the Potts model on bowtie lattices.
 For the case of $q=1$, the critical frontier yields the thresholds of bond percolation on these lattices, which are exactly consistent 
with the results given by Ziff {\it et al} [J. Phys. A {\bf 39}, 15083 (2006)]. 
 For the $q=2$ Potts model on the bowtie-A lattice, the critical point is in agreement with that
of the Ising model on this lattice, which has been exactly solved. 
Furthermore, we do extensive Monte Carlo simulations of Potts model on the bowtie-A lattice with noninteger $q$. Our numerical results, 
which are accurate up to 7 significant digits, are consistent with the theoretical predictions.
We also simulate the site percolation on the bowtie-A lattice, and the threshold is $s_c=0.5479148(7)$. In the simulations of bond percolation and site
percolation, we find that the shape-dependent properties of the percolation model on the bowtie-A lattice are somewhat different from those of an isotropic lattice,
 which may be caused by the anisotropy of the lattice.
\end{abstract}
\maketitle 
\section{Introduction}
 The reduced Hamiltonian of $q$-state Potts model\cite{potts,wfypotts} can be written as 
 \begin{eqnarray}
 -{\mathcal H}/k_BT=K\sum\limits_{<i,j>}\delta_{\sigma_i\sigma_j},\label{pottsmodel}
 \end{eqnarray}
 where $\sigma_i$, $\sigma_j$ are the Potts spins on sites $i$ and $j$. The Potts spin can take $q$ values, namely $\sigma_i=0,1,\cdots,q-1$. 
The sum takes over all nearest-neighboring sites $<i,j>$. $K$ is the coupling constant between spin $\sigma_i$ and $\sigma_j$.
 This model can be mapped to the random-cluster model\cite{FK1,FK2} with partition sum 
 \begin{eqnarray}
 Z_{\rm rc}=\sum\limits_{\mathcal G} v^{N_b} q^{N_c}, \label{rcmodel}
 \end{eqnarray}
where ${\mathcal G}$ is a graph on lattice ${\mathcal L}$ that consists of $N_b$ random occupied bonds and $N_c$ clusters. $v=\exp K-1$ is the bond weight.
$q$ is the cluster weight, which is not restricted to be integer in (\ref{rcmodel}).
 An edge is set to be occupied by a `bond' with probability  $p=v/(1+v)$ or vacant with probability $1-p$. 
A cluster is defined as a group of sites connected by the bonds.

\begin{figure}[htpb]
\includegraphics[scale=0.25]{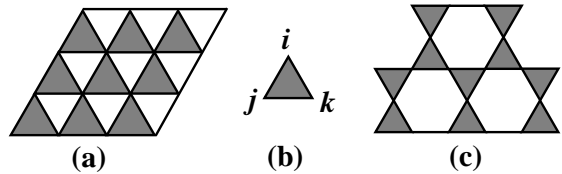}
\caption{(a) the triangular-type lattice, (b) the triangular cell, (c) the kagome-type lattice. The hatched triangular cells can have internal structures.}
\label{tri&kag}
\end{figure}

\begin{figure}[htpb]
\includegraphics[scale=0.16]{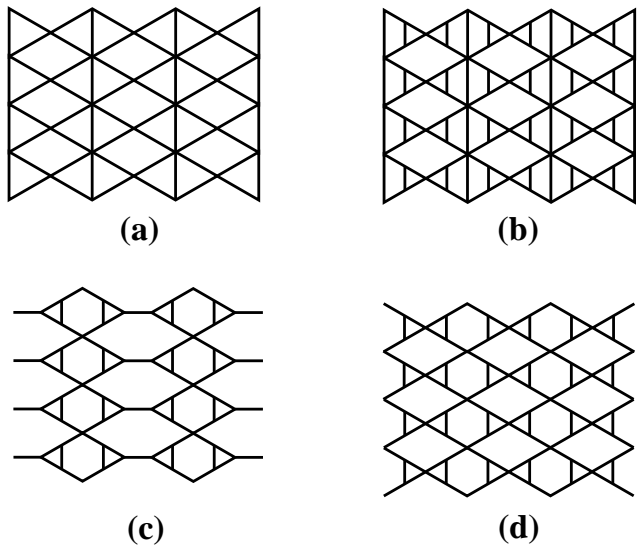}
\caption{Bowtie lattices: (a) bowtie-A lattice, (b) bowtie-B lattice, (c) bowtie-C lattice, (d) bowtie-D lattice.} 
\label{bowtie}
\end{figure}

When $q=1$, the random-cluster model (\ref{rcmodel}) reduces to the bond percolation model\cite{Stauffer}, 
 in which the physicists have special interests.
A central problem of the research in percolation model is the determination of the thresholds of the percolation transition on different lattices. 
After decades of investigation, the percolation thresholds on a lots of lattices have been determined. Most of the results are numerical ones,
exact results are scarce. An important exact result for bond percolation threshold is the one for the triangular-type lattice (see Fig. \ref{tri&kag}(a)).
 As is summarized by Ziff {\it et al.}\cite{Ziffperco,Chayes2006}, 
the percolation threshold of the lattice is determined by 
\begin{eqnarray}
P(i,j,k)=P(\bar{i},\bar{j},\bar{k}), \label{ABC}
\end{eqnarray} 
where $P(i,j,k)$ is the probability that the three apexes $i$, $j$ and $k$ of the triangular cell (Fig. \ref{tri&kag}(b)) are connected in a same cluster
via the bonds in the cell, while 
$P(\bar{i},\bar{j},\bar{k})$ is the probability that the three apexes are isolated in three different clusters. There are other expressions to 
determine the percolation thresholds of the triangular-type lattices\cite{Ziffperco}, but they are in fact equivalent to (\ref{ABC}). 
(\ref{ABC}) was used to determine the exact percolation thresholds of a series of lattices, such as the martini lattices\cite{martinisite,Ziffperco} 
and the stack-of-triangle lattices\cite{voidsperco}. All of these lattices belong to the triangular-type lattice, with different internal structure in the triangular cells. Ziff and Scullard\cite{Ziffexact} extend (\ref{ABC}) to a series of lattices which 
they call bowtie lattices, as shown in Fig. \ref{bowtie}. The threshold of bond percolation  
on the bowtie-A lattice was first determined by Wierman\cite{st-tri2}, using the star-triangle transformation\cite{st-tri1}.

 When $q\ne1$, the random-cluster model (\ref{rcmodel}) can be considered as a general percolation model with cluster weight in the partition sum. 
The determination of the critical points of this model on different lattices is also a challenge in statistical physics. 
 Similar to the problem of percolation threshold, exact critical points of the random-cluster model are also scarce.
 Up to now, exact critical points were found for only a small number of two-dimensional lattices, such as the square lattice, the honeycomb lattice, the martini lattices\cite{pottsmartini}, 
and the stack-of-triangle lattices\cite{paperI,paperII}. In these lattices, the honeycomb lattice and the martini lattices belong to the triangular-type 
lattice, with complex internal structure in the triangular cells.

For the Potts model on the triangular-type lattice, the partition sum can be written as
\begin{eqnarray}
Z_{\rm tri}=\sum\limits_{\{\sigma\}} \prod\limits_{\Delta} W_{\Delta}(i,j,k), \label{Ztri}
\end{eqnarray}
where $i,j,k$ are the three apexes of the cell, and the sum and product take over all the cells. 
$W_{\Delta}(i,j,k)$ is the Boltzman weight of a hatched triangular cell, it can be written in the form of 
\begin{eqnarray}
W_{\Delta}(i,j,k)=A+B(\delta_{\sigma_i\sigma_j}+\delta_{\sigma_j\sigma_k}+\delta_{\sigma_k\sigma_i})+C\delta_{\sigma_i\sigma_j\sigma_k}.
\end{eqnarray}
For example, if the internal structure of the cells are simple triangles, then
\begin{eqnarray}
W_{\Delta}(i,j,k)&=&\exp[K(\delta_{\sigma_i\sigma_j}+\delta_{\sigma_j\sigma_k}+\delta_{\sigma_k\sigma_i})]\nonumber\\
&=&(1+v\delta_{\sigma_i\sigma_j})(1+v\delta_{\sigma_j\sigma_k})(1+v\delta_{\sigma_k\sigma_i})\nonumber\\
&=&1+v(\delta_{\sigma_i\sigma_j}+\delta_{\sigma_j\sigma_k}+\delta_{\sigma_k\sigma_i})+(v^3+3v^2)\delta_{\sigma_i\sigma_j\sigma_k},
\end{eqnarray}
where $v=\exp K-1$ and $\delta_{\sigma_i\sigma_j\sigma_k}=\delta_{\sigma_i\sigma_j}\delta_{\sigma_j\sigma_k}$.
This gives $A=1, B=v$, and $C=v^3+3v^2$. 

 Kelland\cite{Kelland} showed that the partition sum (\ref{Ztri}) is a self-dual one with self-dual point
\begin{eqnarray}
qA=C. \label{qA}
\end{eqnarray}
Wu and Lin\cite{WuandLin} established rigorously that this point is 
indeed the critical frontier in the ferromagnetic regime 
\begin{eqnarray}
3B+C>0,\quad 2B+C>0.
\end{eqnarray}
This is a very important result  in determining the critical point of Potts model. 
For example, substituting the expressions $A=1$ and $C=v^3+3v^2$ to (\ref{qA}), we get the critical frontier of the Potts model on 
the triangular lattice,
which writes $v^3+3v^2=q$. Substituting $p=v/(1+v)$ to this equation with $q=1$, 
we get the threshold of bond percolation on the triangular lattice, i.e., $p_c=2\sin(\pi/18)$, which is a famous result\cite{st-tri1}.

Basing on (\ref{qA}), Wu gives the  critical points
of the kagome-type lattices (Fig. \ref{tri&kag}(c)) through a homogeneous conjecture\cite{paperI}
\begin{eqnarray}
(q^2A+3qB+C)^2-3(qB+C)^2-(q-2)C^2=0.
\end{eqnarray}
 This conjecture was checked by a lot of accurate numerical 
results given by the finite-size scaling analysis\cite{paperII}, which shows that Wu's conjecture was not 
exact but an excellent approximation.

In this paper, we will show that (\ref{qA}) can be extend to the bowtie lattices for determining the exact critical points of 
the random-cluster model on these lattices.
This is very similar to Ziff's extension of (\ref{ABC}) to this type of lattices for determining the bond percolation thresholds. 
Furthermore, we do Monte Carlo simulations of the random-cluster model on the bowtie-A lattice, the numerical results of the critical 
points are consistent with the theoretical predictions.
 In the procedure of simulations, we find that the shape-dependent properties\cite{shapeuniv1,shapeuniv} of the percolation model on the 
bowtie-A lattice are somewhat different from those of an isotropic lattice, by comparing some numerical results of bond percolation and site percolation.

\section{Critical frontier on bowtie lattices}
\label{bowtie_cri}
In fact, (\ref{ABC}) is exactly equivalent to the $q=1$ case of (\ref{qA}). $A$ is the sum of the weights
of sub-graphs with the three apexes $i,j,k$ are isolated in three different clusters, corresponding to the 
probability $P(\bar{i},\bar{j},\bar{k})$. $C$ is the sum of the weights of sub-graphs with the three apexes connected in a same cluster, 
corresponding to the probability $P(i,j,k)$. For example, in a simple triangle, $P(\bar{i},\bar{j},\bar{k})=(1-p)^3$ and $P(i,j,k)=p^3+3p^2(1-p)$.
Substituting $p=v/(1+v)$ to these expressions, one can easily find that $P(\bar{i},\bar{j},\bar{k})$ and  $P(i,j,k)$ are different to $A$ and $C$ respectively with 
a constant $[1/(1+v)]^3$.  This property holds for any triangular cell (but the constant may be different for different internal structure).
In Ref. \cite{paperII}, a computer program was used to calculate the expressions of $A,B,C$ of $n\times n$ stack-of-triangle cell up to $n=4$.
 In Ref. \cite{voidsperco}, Haji-Akbari and Ziff give the expressions of $P(\bar{i}, \bar{j},\bar{k})$ and $P(i,j,k)$. Comparing these
expressions, we can see more clearly the equivalence of (\ref{ABC}) and the $q=1$ case of (\ref{qA}).

\begin{figure}[htpb]
\includegraphics[scale=0.20]{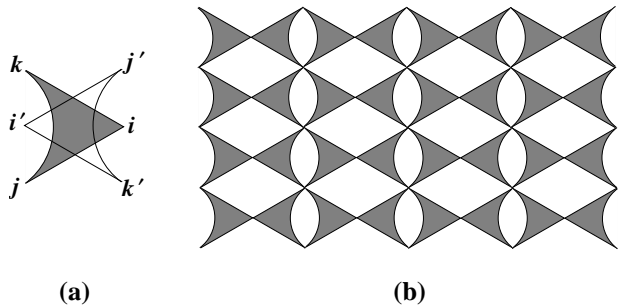}
\caption{(a) duality transformation of a triangular cell, (b) the bowtie-generating lattice, which is self-dual under the duality transformations of 
the cells. Four bowtie lattices as shown in Fig. \ref{bowtie} can be generated by this lattice by inserting different internal structures in the 
hatched triangular cells.} 
\label{bowtie_generate}
\end{figure}

\begin{figure}[htpb]
\includegraphics[scale=0.20]{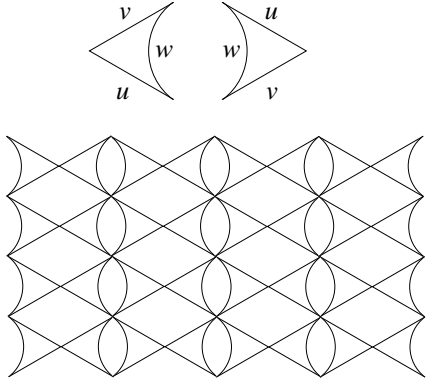}
\caption{A lattice constructed by inserting simple triangles in the cells of bowtie-generating lattice and its bond weights of the random-cluster model (top).} 
\label{pre_bA}
\end{figure}

(\ref{ABC}) follows the duality arguments\cite{dual0,dual}, it
can be used to calculate the threshold of bond percolation on a lattice composed of triangular cells, and the lattice must
be self-dual under the transformation shown in Fig. \ref{bowtie_generate}(a).  The triangular-type lattice obviously meets this condition. Another type of lattice 
that meets this condition is the lattice shown in Fig. \ref{bowtie_generate}(b), and the four bowtie lattices can be generated by inserting different internal 
structures in the hatched triangular cells of this lattice. This lattice is first given by Ziff and Scullard\cite{Ziffexact}, we call this lattice the 
`bowtie-generating lattice', because it `generates' the four bowtie lattices. If simple triangles are inserted in the cells of the bowtie-generating lattice, it
generates a lattice with double bonds as shown in Fig. \ref{pre_bA}. The bond percolation with uniform probability $p$ on the bowtie-A lattice can be got by
 setting the double bonds with probability $1-\sqrt{1-p}$, which is the Wierman's trick\cite{st-tri2}.

Similar to (\ref{ABC}), (\ref{qA}) also follows the duality arguments\cite{Kelland, WuandLin}, this inspires us to extend (\ref{qA}) to calculate the 
critical points of random-cluster model on other lattice which meets the condition of self-duality. As an example, we calculate the critical points of 
the bowtie lattices in this paper.

\subsection{bowtie-A lattice}
For generality, we set the coupling constants of the three bonds of the triangle in the double-bond lattice with $K,M,N$. The bond weights
 are $v=\exp{K}-1, u=\exp{M}-1$ and $w=\exp{N}-1$, which are shown in Fig. \ref{pre_bA}. The expressions of $A$ and $C$ are
\begin{eqnarray} 
A&=&1,\\
C&=&uvw+uv+vw+wu.
\end{eqnarray}
Substituting $A$ and $C$ to (\ref{qA}), we get the critical frontier of the Potts model on the double-bond lattice
\begin{eqnarray}
uvw+uv+vw+wu=q. \label{double_bond}
\end{eqnarray}
Setting 
\begin{eqnarray}
M=K, N=K/2,
\end{eqnarray}
namely
\begin{eqnarray}
u=v, w=\sqrt{1+v}-1,\label{uw}
\end{eqnarray}
we get a Potts model with uniform coupling constant $K$ or random-cluster model with uniform bond 
weight $v$ on the bowtie-A lattice. Substituting (\ref{uw}) to (\ref{double_bond}), we get the critical 
frontier of the Potts model on the bowtie-A lattice, which writes
\begin{eqnarray}
(v^2+2v)(\sqrt{1+v}-1)+v^2=q, \quad \mbox{bowtie-A lattice.}\label{bowtieAcri}
\end{eqnarray}
Solving (\ref{bowtieAcri}) with $q=1$, we get $v_c=0.679~312~786$ or $K_c=0.518~384~653$. This yields the threshold of bond percolation on the 
bowtie-A lattice, i.e., $p_c=v_c/(1+v_c)=1-\exp(-K_c)=0.404~518~319$, which is exactly equal to the result that given by Wierman\cite{st-tri2} 
or Ziff {\it et al.}\cite{Ziffexact}.  

The case of $q=2$, which is the Ising model, has been exactly solved by Str$\check{\rm e}$cka and $\check{\rm C}$anov$\acute{\rm a}$\cite{bowtieIsing} through 
mapping the model into a free-fermion eight-vertex model, and the critical points is given by 
\begin{eqnarray}
[\cosh(4K_I)-1]e^{2K_I}=2, \label{bAIsing}
\end{eqnarray}
where $K_I=K/2$ is the coupling constant of nearest-neighboring Ising spins. Solving (\ref{bAIsing}), we get $K_{Ic}=0.333~135~959$.
Furthermore, if we solve (\ref{bowtieAcri}) with $q=2$, we get $K_c=0.666~271~918$. 
The two results are exactly consistent with each other.

The cases of $q=1$ and 2 show the correctness of (\ref{bowtieAcri}) for the bowtie-A lattice. For the other case, especially the case that $q$
 is not an integer, the correctness of (\ref{bowtieAcri}) is not verified. Thus we do extensive Monte Carlo simulation and finite-size scaling analysis 
for the Potts model on the bowtie-A lattice, the numerical results will be given in Sec. \ref{MC}.

Setting the double-bond weights $w=0$ (namely $N=0$), (\ref{double_bond}) gives the critical frontier 
of Potts model on the square lattice
\begin{eqnarray}
uv=q,
\end{eqnarray}
which is a famous result\cite{Baxter}.

\subsection{bowtie-B, and bowtie-D lattices}
The critical frontier of the Potts model on the bowtie-B, the bowtie-C and the bowtie-D lattices can be obtained by the same way that we used for the bowtie-A lattice.
In this paper, we only pay attention to the random-cluster model on these lattices with uniform bond weight $v$. 
\begin{figure}[htpb]
\includegraphics[scale=0.35]{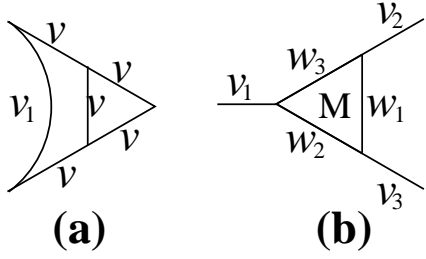}
\caption{(a) internal structure for constructing the bowtie-B and bowtie-D lattices, (b) internal structure for constructing the lattice as shown 
in Fig. \ref{bowtiec}(a). $v=\exp K-1, v_i=\exp K_i -1, w_i=\exp L_i-1$ are bond weights, $M$ is the 3-site coupling
 constant between the three spins of the triangle in center.} 
\label{subnets}
\end{figure}

In order to get the critical frontier of the Potts model on the bowtie-B and the bowtie-D lattices, we insert the network as shown in Fig. \ref{subnets}(a) to the
 cells of the bowtie-generating lattice, this generates a lattice that also has double bonds. The expression of $A$ and $C$ are 
\begin{eqnarray}
 A&=&q^2+5qv+8v^2+v^3,\\
 C&=&3v^4+v^5+v_1(2qv^2+8v^3+5v^4+v^5),
\end{eqnarray}
where $v=\exp K-1$ and $v_1=\exp K_1-1$ are the bond weights (see Fig. \ref{subnets}(a)). Substituting them to (\ref{qA}), we get the critical frontier 
\begin{eqnarray}
 3v^4+v^5+v_1(2qv^2+8v^3+5v^4+v^5)-q(q^2+5qv+8v^2+v^3)=0.
\end{eqnarray}
Setting $K_1=K/2$, namely $v_1=\sqrt{1+v}-1$, we get the critical frontier of the Potts model on the bowtie-B lattice with uniform coupling constant $K$
\begin{eqnarray}
3v^4+v^5+(\sqrt{1+v}-1)(2qv^2+8v^3+5v^4+v^5)-q(q^2+5qv+8v^2+v^3)=0,&&\nonumber\\
 \mbox{bowtie-B lattice.}&& \label{bowtieB}
\end{eqnarray}
Solving this formula with $q=1$, we get the critical point $v_c=1.142~305~296$ or $K_c=0.761~882~490$. This yields the threshold of bond percolation 
$p_c=0.533~213~122$, which is exactly consistent with the value given in Ref. \cite{Ziffexact}. 

Setting $K=0$, namely $v_1=0$, we get the critical frontier 
of the Potts model on the bowtie-D lattice with uniform $K$
\begin{eqnarray}
 3v^4+v^5-q(q^2+5qv+8v^2+v^3)=0,
\quad \mbox{bowtie-D lattice.} \label{bowtieD}
\end{eqnarray}
Solving this formula with $q=1$, we get the critical point $v_c=1.669~919~079$ or $K_c=0.982~048~164$. This yields the bond percolation threshold
$p_c=0.625~456~813$, which is exactly consistent with the value given in Ref. \cite{Ziffexact}. (\ref{bowtieD}) is also the critical frontier of the Potts model 
on the martini-A lattice, see Ref. \cite{pottsmartini} for details.

\subsection{bowtie-C lattice}
\begin{figure}[htbp]
\includegraphics[scale=0.16]{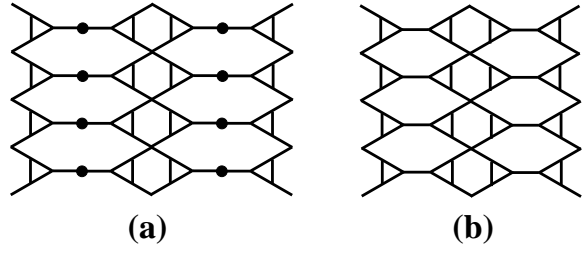}
\caption{After the decimation of the dotted spins, the lattice (a) becomes the lattice (b), namely the bowtie-C lattice.}
\label{bowtiec}
\end{figure}
By inserting the network as shown in Fig. \ref{subnets}(b) to the triangular cells of the bowtie-generating lattice,
 we get a lattice as shown in Fig. \ref{bowtiec}(a). The expressions of $A, C$ for the network as shown in Fig. \ref{subnets}(b) have been 
given by Wu\cite{pottsmartini}:
\begin{eqnarray}
A&=&v_1v_2v_3+v_1v_2(q+w_1+w_2)+v_2v_3(q+w_2+w_3)+v_3v_1(q+w_3+w_1)\nonumber\\
&&+(q+v_1+v_2+v_3)\times [q^2+q(w_1+w_2+w_3)+h]\\
C&=&v_1v_2v_3h
\end{eqnarray}
where 
\begin{eqnarray}
v_i&=&e^{K_i} -1, w_i=e^{L_i} -1\nonumber\\
h&=&e^{M+L_1+L_2+L_3}-e^{L_1}-e^{L_2}-e^{L_3}+2.
\end{eqnarray}
$K_i,L_i$ are the 2-site coupling constant, $M$ is the 3-site coupling constant.
Setting the bond weights $w_1=w_2=w_3=v_2=v_3=v$ (i.e., $L_1=L_2=L_3=K_2=K_3=K$) and the 3-site coupling constant $M=0$,
we get the critical frontier of the Potts model on the lattice as shown in Fig. \ref{bowtiec}(a):
\begin{eqnarray}
q(q+2v)[q^2+3qv+v^2(4+v)]+[q^3+5q^2v-v^4(3+v)+qv^2(8+v)]v_1=0. \label{bowtieC_cri}
\end{eqnarray}
Decimating the dotted spins in the lattice as shown in Fig. \ref{bowtiec}(a), we get the lattice as shown in Fig. \ref{bowtiec}(b), which is 
the bowtie-C lattice. If the bond weight $v_1$ is set as $v_1=v+\sqrt{v^2+vq}$, we get the critical frontier of the Potts model on the 
bowtie-C lattice with uniform bond weight $v$:
\begin{eqnarray}
q(q+2v)[q^2+3qv+v^2(4+v)]+[q^3+5q^2v-v^4(3+v)+qv^2(8+v)](v+\sqrt{v^2+qv})=0. &&\nonumber\\
\mbox{bowtie-C lattice.}\quad&& \label{bowtieC}
\end{eqnarray}
Solving this formula with $q=1$, we get the critical point $v_c=2.057~439~254$ or $K_c=1.117~577~720$. This yields the bond percolation threshold
$p_c=0.672~928~906$, which is exactly consistent with the result that given in Ref. \cite{Ziffexact}.

We also solve (\ref{bowtieAcri}), (\ref{bowtieB}), (\ref{bowtieD}), and (\ref{bowtieC}) with other value of $q$, some results are listed in Table. \ref{bowtie_exact}.
In the table, we only give the results with $q\le 4$, because the phase transition of the Potts model becomes discontinuous when $q>4$\cite{Baxter}.

\begin{table}[htbp]
\caption{Exact critical points of the Potts model on the bowtie lattices, A = bowtie-A lattice, B = bowtie-B lattice, C = bowtie-C lattice,
D = bowtie-D lattice.}
 \begin{tabular}{c|c|c|c|c}
    \hline
      $q$   &$K_c$(A)     &$K_c$(B)    &$K_c$(C) &$K_c$(D)\\ 
    \hline
    $1.0$   &0.518 384 653 &0.761 882 490 &1.117 577 720 &0.982 048 164           \\
    $1.5$   &0.601 796 427 &0.880 501 865 &1.263 365 870 &1.116 885 906 \\
    $2.0$   &0.666 271 918 &0.971 526 940 &1.372 908 989 &1.218 755 726 \\
    $2.5$   &0.719 275 592 &1.045 947 678 &1.461 235 663 &1.301 189 204 \\
    $3.0$   &0.764 490 171 &1.109 155 997 &1.535 511 667 &1.370 686 799  \\
    $3.5$   &0.804 030 933 &1.164 234 324 &1.599 748 425 &1.430 907 221  \\
    $4.0$   &0.839 235 249 &1.213 123 525 &1.656 430 741 &1.484 126 262 \\
    \hline
\end{tabular}
\label{bowtie_exact}
\end{table}

\subsection{bowtie-dual lattices}
The critical frontier of the Potts model on the dual of the bowtie lattice, namely the bowtie-dual lattice (see Fig. 5 of Ref. \cite{Ziffexact}),
 can be obtained by the dual relation\cite{dual}
\begin{eqnarray}
v^*\cdot v=q,
\end{eqnarray}
or 
\begin{eqnarray}
(e^{K^*}-1)(e^K-1)=q,
\end{eqnarray}
where $K$ is the coupling constant on the bowtie lattice, and $K^*$ is the coupling constant on the dual lattice. $v=\exp K-1$ and $v^*=\exp K^*-1$
 are the bond weights on the bowtie lattice and the bowtie-dual lattice, respectively. 
Thus the critical frontier of the Potts model on the bowtie-dual lattice is given by 
\begin{eqnarray}
(e^{K^*_c}-1)(e^{K_c}-1)=q.
\end{eqnarray}

\section{Monte Carlo simulation and numerical results}
\label{MC}
\subsection{Simulation of the Potts model on the bowtie-A lattice}
Using the cluster algorithm\cite{Qian}, we do Monte Carlo simulation of the Potts model on the bowtie-A lattice with $q\ge1$. 
This cluster algorithm is a combination of Swendsen-Wang algorithm\cite{SW} and `coloring trick'\cite{color1,color2}, which drastically reduces
 the critical slowing down problem.
This type of algorithm has been developed and used to simulate O($n$) loop model\cite{DengAlg,Ding2009}, Eulerian bond-cubic model\cite{DingCubic} and so forth.

In the random-cluster model (\ref{rcmodel}), the behaviors of the clusters are similar to those of percolation model.  
There are small clusters on the configurations when the bond weight $v$ is small, but the clusters will growth as the value of $v$ increases. 
The largest cluster diverges at the critical point $v_c$ (or $K_c$) in the thermodynamic limit, which is called a `percolating cluster'. 
However, for a finite system, the largest cluster may be different to the `percolating cluster'. 
There are different rules to define the `percolating cluster' for a finite system with periodic boundary condition. Here we use 
the `wrapping cluster'\cite{wrapping}, which
is defined as a cluster that connects itself along at least one of the periodic directions. 

Basing on the definition of the `wrapping cluster', the `wrapping probabilities' on a rectangular lattice are defined as 
\begin{eqnarray}
R_x&=&\langle \mathcal{R}_x \rangle,\nonumber\\
R_y&=&\langle \mathcal{R}_y \rangle, \nonumber\\
R_{xy}&=&\langle \mathcal{R}_{xy}\rangle,
\end{eqnarray}
where $\langle \cdots \rangle$ stands for ensemble average. $\mathcal{R}_x$ is 1 (zero) if there is a (no) cluster that wraps along the $x$ direction, 
 whether or not the cluster wraps along $y$ direction.
The definition of $\mathcal{R}_y$ is similar to $\mathcal{R}_x$. $\mathcal{R}_{xy}$ is defined as 
$\mathcal{R}_{xy}=\mathcal{R}_x\cdot\mathcal{R}_y$, whose value is 1 if and only if there exists a cluster that wraps along $x$ and $y$ directions at the same time.
\begin{figure}[htbp]
\includegraphics[scale=0.2]{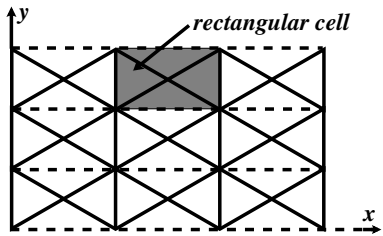}
\caption{Definition of wrapping probabilities on the bowtie-A lattice.}
\label{bowtie_R}
\end{figure}

The bowtie-A lattice that we simulated in the paper can be viewed as a `rectangular lattice' with internal structure in the rectangular cells, 
see Fig. \ref{bowtie_R}. This 
rectangular lattice is obviously not symmetry between $x$ and $y$ directions; therefore the values of $R_x$ and $R_y$ will be not the same, 
which will be shown concretely later.
The wrapping probabilities $R_x$, $R_y$ and $R_{xy}$ have the similar finite-size scaling behavior\cite{fssa,fssb}
\begin{eqnarray}
R=R_0+a_1(K-K_c)L^{y_t}+a_2(K-K_c)L^{2y_t}+\cdots+b_1L^{y_1}+b_2L^{y_2}+\cdots, \label{Rfss}
\end{eqnarray}
where $R$=$R_x$, $R_y$ or $R_{xy}$. $R_0$ is the value of $R$ at the critical point $K_c$. In present paper, we call it `critical wrapping probability'.
$y_t$ is the thermal exponent of the Potts model, and $y_i$ the correction-to-scaling exponent, which has negative value. $a_i$ and $b_i$ 
are unknown parameters. The value of $y_t$ can be derived by Coulomb gas method\cite{CG} or conformal invariance\cite{conformal}
\begin{eqnarray}
 y_t&=&3-\frac{3}{2g}, \label{yt}\\
{\rm with} \quad \sqrt{q}&=&-2\cos(\pi g), \quad 1/2\le g \le 1,
\end{eqnarray}
where $g$ is the coupling constant of the Coulomb gas. 

\begin{figure}[htpb]
\includegraphics[scale=1.0]{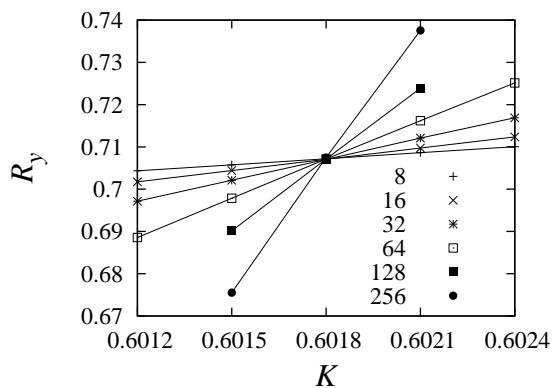}
\caption{$R_y$ versus $K$ for various system size of $q=1.5$ Potts model. All the error bars are much smaller than the data points, the lines between points are added for illustration purpose.} 
\label{Ry15}
\end{figure}

In order to illustrate our numerical procedure, we take the $q=1.5$ Potts model as an example. 
The cluster algorithm easily allows us to do meaningful simulations for system
 with  linear size up to $L=256$. After equilibrating the system, $10^9$ samples were taken for each
value of $K$ for $8\le L\le 64$, and $1.4\times 10^8$ samples for $128\le L\le 256$.
 Figure. \ref{Ry15} is an illustration of $R_y$ versus $K$ with different $L$ for $q=1.5$ Potts model.
Using the Levenberg-Marquardt least squares algorithm, we fit the data according to the finite-size-scaling formula (\ref{Rfss}). The fitting yields 
$R_{0y}=0.7070(2)$, $R_{0x}=0.4363(2)$, $R_{0xy}=0.3780(1)$, $K_c=0.6017964(6)$ and $y_t=0.887(1)$. In these results, 
we can see that the critical point $K_c$ is consistent with our theoretical
prediction $K_c=0.60179643$, and the thermal exponent is consistent with the Coulomb gas prediction $y_t=0.8867$. 

We also simulate the cases of $q=1, 2.5, 3, 3.5$ and 4.  All the numerical results and the theoretical predictions  
are summarized in Table. \ref{bowtieA_num}. 
In the fitting of the data for $q=4$ Potts model, logarithmic corrections\cite{log,log1,log2} should be included. Instead of (\ref{Rfss}), we fit the data
according to 
\begin{eqnarray}
R=R_0+a_1(K-K_c)L^{y_t}(\log L)^{y_1}+a_2(K-K_c)^2L^{2y_t}(\log L)^{2y_1}+\cdots+b_1\frac{\log \log L}{\log L}+\frac{b_2}{\log L},
\end{eqnarray}
with $y_1<0$. However, we can see (from the table) that the 
results for $q=4$ Potts model are relatively inaccurate. This is because of the limited system size.
For an accurate fitting with logarithmic corrections,  data with larger system sizes are necessary. 

\begin{table}[htbp]
\caption{Critical properties of the Potts model on the bowtie-A lattice, T=Theoretical results, N=Numerical results. The value of $K_c$ (N) can be obtained by fitting 
the data of $R_y$, $R_x$, or $R_{xy}$, here we list the best estimation.}
 \begin{tabular}{c|l|l|l|l|l|l|l}
\hline
     $q$      & $K_c$ (T)  & $K_c$ (N)    &$y_t$ (T) & $y_t$ (N) & $R_{0y}$ & $R_{0x}$ & $R_{0xy}$\\
    \hline
    $1$     &0.518 384 653 &0.5183847(7)  &0.7500    &0.750(1)   &0.6490(1)&0.3838(1)&0.3161(1)  \\
    $1.5$   &0.601 796 427 &0.6017964(6)  &0.8867    &0.887(1)   &0.7070(2)&0.4363(2)&0.3780(1)\\
    $2.5$   &0.719 275 592 &0.7192756(11) &1.1018    &1.101(2)   &0.7750(1)&0.5151(2)&0.4729(2)\\
    $3$     &0.764 490 171 &0.7644906(9)  &1.2000    &1.197(6)   &0.7975(1)&0.550(1) &0.515(1)\\
    $3.5$   &0.804 030 933 &0.8040314(25) &1.3050    &1.296(12)  &0.8156(2)&0.585(1) &0.557(1)\\
    $4$     &0.839 235 249 &0.839235(2)   &1.5000    &1.50(7)    &0.829(5)  &0.62(1) &0.60(1)\\
    \hline
\end{tabular}
\label{bowtieA_num}
\end{table}

\subsection{Simulation of site percolation}
We also simulate the site percolation on the bowtie-A lattice, the finite-size scaling analysis gives $s_c=0.5479148(7)$, 
$y_t=0.750(1)$, $R_{0y}=0.6090(1)$, $R_{0x}=0.4286(1)$ and  $R_{0xy}=0.3349(1)$.
 The value of $s_c$ coincides with 0.5475(8), given by van der Marck\cite{Marck}, and our result is much more accurate.

\section{Shape-dependent properties of the bowtie-A lattice}
We also do Monte Carlo simulations of the bond and site percolation models 
on the square lattice. For the percolation thresholds, the numerical estimations give $p_c=0.5000000(4)$ (bond), and $s_c=0.5927460(5)$ (site). The accuracy 
of $s_c$ is somewhat lower than the result given by Feng {\it et al}\cite{Feng}. The numerical results of the critical wrapping probabilities are listed in 
Table \ref{bowtieA_shapeuniv}, these results give $R_{0e}$=$R_{0x}+R_{0y}-R_{0xy}$=0.6902(5) for the bond percolation and 0.6905(5) for the site percolation, 
which are consistent with the literature value $R_{0e}=0.6904737$\cite{wrapvalue,shapeuniv}. $R_{0e}$ is the critical value of the probability that
there is a cluster wrapping along one or both of the two coordinate directions.

\begin{figure}[htpb]
\includegraphics[scale=0.25]{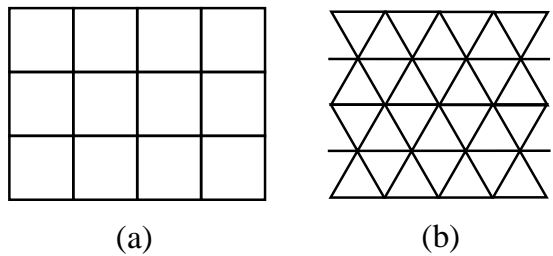}
\caption{(a) A rectangular lattice (on macroscopic level) with square symmetry on microscopic level, with aspect ratio $r=3/4$;
 (b) A rectangular lattice (on macroscopic level) with triangular symmetry on microscopic level, with aspect ratio $r=\sqrt{3}/2$.}
\label{shape}
\end{figure}

For a lattice that is isotropic on microscopic level (equivalent in the coordinate directions, or with the triangular or hexagonal symmetry, etc.), 
the values of the wrapping probabilities are functions of the system shape and boundary conditions\cite{shapeuniv,wrapvalue}, which are independent of the
percolation type.
In saying the shape, it is on the macroscopic level. For example, a macroscopic rectangular lattice may be a lattice with square or triangular symmetry on 
microscopic level, see Fig. \ref{shape} for examples or see Ref. \onlinecite{shapeuniv1,shapeuniv} for details. 
From now on, in saying `isotropic' or `anisotropic' in the text, it is on the microscopic level.
For an isotropic rectangular lattice with periodic boundary conditions (not twisted), 
the value of the wrapping probability $R_{0e}$ is determined by the aspect ratio of the lattice\cite{shapeuniv}
\begin{eqnarray}
R_{0e}(r)=1-\frac{1}{2}[Z_c(8/3,r)-Z_c(2/3,r)], \label{R0}
\end{eqnarray}
where $r$ is the aspect ratio and $Z_c(h,r)$ is
\begin{eqnarray}
Z_c(h)=\frac{\sqrt{h/r}}{\eta^2(w)} \sum\limits_{n=-\infty}^{\infty}\sum\limits_{n^\prime=-\infty}^\infty \exp \big\{-\frac{\pi h}{r}[n^{\prime2}+n^2r^2]  \big\},
\end{eqnarray}
$\eta(w)=w^{1/24}\prod\limits_{k=1}^{\infty}(1-w^k)$ is the Dedekind eta function and $w=e^{-2\pi r}$. When $r=1$, (\ref{R0}) gives $R_{0e}=0.6904737$ for the square lattice.

In our Monte Carlo simulations, the bowtie-A lattice is viewed as a rectangular lattice (with periodic b.c.), as shown in Fig. \ref{bowtie_R}, 
and the aspect ratio is $r=1/\sqrt{3}$. However,
the wrapping probability of the bowtie-A lattice is not determined by this aspect ratio. 
(\ref{R0}) gives $R_{0e}(1/\sqrt{3})=0.7439918$, this is inconsistent with 
our numerical values $R_{0e}=0.7167(1)$ (bond) and 0.7027(1) (site). Furthermore, 
we can see that the values of $R_{0}$ (include $R_{0e}, R_{0x}, R_{0y}$, and $R_{0xy}$, see Table. \ref{bowtieA_shapeuniv}.) 
for the bond percolation are different from those for the site percolation. These results show the difference between the bowtie-A lattice 
and an isotropic lattice in the aspect of shape-dependent properties.

The difference may be caused by the anisotropy of the bowtie-A lattice,
 and (\ref{R0}) is only valid for an isotropic lattice. 
 After mapping the percolation model on the bowtie-A lattice onto the  Gaussian model\cite{wrapvalue},  it will be an anisotropic one. 
Thus a rescaling on $x$ or $y$ direction is required  in order to obtain an
isotropic Gaussian model. After the rescaling, the isotropic model would have an aspect ratio that is not equal to $1/\sqrt{3}$ (Unfortunately, we 
don't know the value of the effective aspect ratio). This is the reason why our numerical value of $R_{0e}$ for the bowtie-A lattice 
doesn't coincide with $R_{0e}(1/\sqrt{3})$.

More important, there is no reason why the bond and site percolation models on the bowtie-A lattice should map onto the same 
(anisotropic) Gaussian model, and after rescaling to the isotropic Gaussian model, there would be different aspect ratios for 
the bond and site percolation models. Thus, there would be different values of the wrapping probabilities for the bond and site percolation models,
 which are verified by our numerical results.


Another quantity that we simulate for investigating the shape-dependent properties is the average 
density of clusters, which is defined as the average number of clusters per site.
We find that on such an anisotropic lattice, the average density of clusters also behaves as\cite{shapeuniv1}
\begin{eqnarray}
n=n_c+b/N+\cdots,
\end{eqnarray}
when the system is at the critical point. $n_c$ is the value of $n$ for the infinite system, and $N$ the total number of sites of a finite system.
$b$ is the excess number of clusters over the bulk value of clusters $n_cN$. For an isotropic lattice with periodic boundary conditions (not twisted), 
the value of $b$ is also a function of system 
shape, which has the same value for the bond and site percolation models\cite{shapeuniv1} 
\begin{eqnarray}
b(r)=\frac{5\sqrt{3}r}{24}+w^{5/4}(2\sqrt{3}r-\frac{1}{2})+w^2(\sqrt{3}r-1)+w^{5/48}+2w^{53/48}-w^{23/16}+w^{77/48}+\cdots,
\end{eqnarray}
where $w=e^{-2\pi r}$. This equation gives $b(1/\sqrt{3})=0.943713$, which is obviously inconsistent with our numerical results for the bowtie-A lattice:
 $b^{\rm B}=0.9120(6)$ and $b^{\rm S}=0.8957(4)$ (B denotes bond percolation, S denotes site percolation).  
More important, the value of $b^{\rm B}$ is different from that of $b^{\rm S}$.
The physical essence of these results for $b$ is the same to that for the wrapping probabilities, it may also be caused by the anisotropy of the bowtie-A lattice.

In conclusion (Table \ref{bowtieA_shapeuniv}), the shape-dependent properties of the percolation model on the bowtie-A lattice are somewhat different from
those of an isotropic lattice, this may be caused by the anisotropy of the bowtie-A lattice.
The bond and site percolation models on the bowtie-A lattice correspond 
with different anisotropies after mapping onto Gaussian models, and thus with different aspect ratios after rescaling to isotropic models,  
which lead to the difference of shape-dependent properties between the bond and site percolation models. 

\begin{table}[htbp]
\caption{Critical values of the wrapping probability $R_0$, the average density of clusters $n_c$ and the 
excess number of clusters $b$ on the square and bowtie-A lattices, Site=Site percolation, Bond=Bond percolation. $^c$=Reference\cite{shapeuniv1},
$^d$=Reference\cite{shapeuniv},$^e$=Reference\cite{wrapvalue},$^f$=Reference\cite{Kleban}}
 \begin{tabular}{l|l|l|l|l|l|l}
\hline
     System           & $R_{0y}$  & $R_{0x}$ &$R_{0xy}$& $R_{0e}$ &$b$  & $n_c$                      \\
    \hline
    Bond (square)     &0.5208(5)  &0.5210(5) &0.3516(5) &0.6902(5) &0.883(1)&0.098076(1) \\
                      &0.5210583$^{d,e}$  &0.5210583$^{d,e}$&0.3516429$^{d,e}$&0.6904737$^{d,e}$&0.8838(5)$^c$&0.0980763(8)$^c$ \\
                      & & & &&0.883756 $^f$ & $\frac{3\sqrt{3}-5}{2}$$^c$\\
    Site (square)     &0.5211(5)(1)&0.5209(5)&0.3514(4) &0.6905(5)&0.8834(11)&0.027598(1)\\
                      &0.5210583$^{d,e}$&0.5210583$^{d,e}$&0.3516429$^{d,e}$&0.6904737$^{d,e}$&0.8832(3)$^c$&0.0275981(3)$^c$ \\
    Bond (bowtie-A)   &0.6490(1)  &0.3838(1) &0.3161(1) &0.7167(1) &0.9120(6)&0.119212(1)  \\
    Site (bowtie-A)   &0.6090(1)  &0.4286(1) &0.3349(1) &0.7027(1) &0.8957(4)&0.023990(1)  \\
    \hline
\end{tabular}
\label{bowtieA_shapeuniv}
\end{table}

\section{Conclusion}
\label{con}

In conclusion, we have given the exact critical frontier of the Potts model on the bowtie lattices.
For $q=1$, the critical frontier yields the thresholds of bond percolation on these lattice, which are consistent with the results given by 
Ziff {\it et al}. For the $q=2$ Potts model on the bowtie-A lattice, the critical point coincides with the exactly solved critical point 
of the Ising model on this lattice.
Furthermore, we do Monte Carlo simulations of the Potts model on the bowtie-A lattice, the numerical results, which includes the cases with 
noninteger $q$, are in agreement with the exact critical points in a high precision.

Furthermore, by comparing the numerical results of the critical wrapping probability $R_0$ and the excess number of clusters $b$ for the 
bond and site percolation models, we find that the shape-dependent properties of the percolation model on the bowtie-A lattice are somewhat different from 
those of an isotropic lattice, which may be caused by the anisotropy of the bowtie-A lattice.
\acknowledgments
We are much indebted to Prof. R. M. Ziff for valuable discussions. We thanks Prof. Wenan Guo for a critical reading of the manuscript.
 This research is supported by the National Science Foundation of
China  under Grant \#11175018, and by the High Performance Scientific Computing Center (HSCC) of the Beijing Normal University.

\end{document}